\documentstyle[12pt]{article}
\begin{document}
\begin{center}
{\Large \bf On quantization of gravity}
\bigskip

{\large D.L.~Khokhlov}
\smallskip

{\it Sumy State University, R.-Korsakov St. 2\\
Sumy 244007 Ukraine\\
e-mail: others@monolog.sumy.ua}
\end{center}

\begin{abstract}
The theory of gravity without free gravitational fields is
considered. It is assumed that gravitational radiation is
some hypothetical matter fields.
Gravitational emission is
a process of the decay of proton into hypothetical
matter fields at the Planck scale.
\end{abstract}
\bigskip

In the Einstein theory of gravity~\cite{Buch} with the Lagrangian
\begin{equation}
L=-{1\over{16\pi G}}\sqrt{g}R,\label{eq:a}
\end{equation}
gravitational field is defined by the components of the metric
tensor $g_{ik}$.
Quantization of gravity reduces to quantization
of the fields $g_{ik}$, with the free gravitational fields
are described by the Einstein equations
\begin{equation}
R_{ik}=0.
\label{eq:ee}
\end{equation}

According to Rosenfeld \cite{Ros2},
the necessity to quantize gravity is an empiric problem.
The mental experiment of Bohr and Rosenfeld
\cite{Ros2}
on measurement of the field values with the use of the
Einstein equivalence principle gives the uncertainty of
measurement of gravitational fields and distances
\begin{equation}
\Delta g(\Delta l)^2\geq\frac{2\hbar G}{c^3}=2l_{Pl}
\label{eq:dg}
\end{equation}
where $l_{Pl}$ is the Planck length.
On the other hand,
uncertainty relation (\ref{eq:dg})
holds in quantum gravity as a commutational relation
\cite{Ros1}.
This means~\cite{Tr} that all the quantum gravitational effects
are less than the uncertainty of measurement in accordance with
(\ref{eq:dg}).
That is quantum and classical gravity are the same,
since we cannot distingwish classical
and quantum gravity in experimental way.
From this it follows that
free gravitational fields given by (\ref{eq:ee})
do not exist.

Let us consider the theory in which gravity exists
only if matter exists
\begin{equation}
G_{ik}=T_{ik}.
\label{eq:mee}
\end{equation}
Free gravitational fields given by (\ref{eq:ee})
do not exist, and gravity is not quantized.
Let us assume that gravitational radiation
is some matter field. Let us take this as
hypothetical massless fermions $\psi_{hyp}$
which do not take part in electromagnetic, weak and strong
interactions.
Let us consider the decay of proton into
hypothetical fermions $\psi_{hyp}$ at the Planck scale
$m_{Pl}=(\hbar c/G)^{1/2}$
\begin{equation}
p\rightarrow \psi_{hyp}\psi_{hyp}.
\label{eq:dp}
\end{equation}
We can use the standard formalism of quantum field
theory~\cite{Bog}.
Take the Lagrangian of the proton decay in the form
\begin{equation}
L(x)\rightarrow{g^2\over m_{Pl}^2}J_{\nu}(x)J^{\nu}(x)
\label{eq:lpd}
\end{equation}
where the charge is given by
\begin{equation}
g=(\hbar c)^{1/2}.
\label{eq:ech}
\end{equation}
The matrix element of the Lagrangian (\ref{eq:lpd})
is of order
\begin{equation}
M\sim{g^2\over m_{Pl}^2}=G.
\label{eq:mel}
\end{equation}
That is the decay of proton into
hypothetical fermions $\psi_{hyp}$
is governed by the Newton constant.
Therefore we can identify the decay of proton into
hypothetical fermions $\psi_{hyp}$ with the process of
gravitational emission.

The probability of the proton decay given by the matrix element
(\ref{eq:mel}) is of order
\begin{equation}
w \sim \frac{c^2}{\hbar}\frac{m_p^5}{m_{Pl}^4}=
\frac{1}{t_{Pl}}\left(\frac{m_p}{m_{Pl}}\right)^5
\label{eq:pdq}
\end{equation}
where $m_{p}$ is the mass of proton,
$t_{Pl}=(\hbar G/c^5)^{1/2}$ is the Planck time.
Let us estimate the probability of proton decay
within the framework of semiclassical theory with the use
of the Einstein formula for gravitational emission~\cite{Lan}
\begin{equation}
-{{dE}\over{dt}}={G\over{45c^5}}{\stackrel\dots D^2}
\label{eq:gem}
\end{equation}
Nonlocalizability of gravitational fields in accordance with
(\ref{eq:dg}) defines an effective inertia moment of proton
\begin{equation}
I\sim m_{Pl}l_{Pl}^2.
\label{eq:imo}
\end{equation}
Due to the presence of spin
proton can be considered as an asymmetric rotator.
The intensity of gravitational emission for the asymmetric
rotator
is given by
\begin{equation}
-{{dE}\over{dt}}\sim
\frac{G}{c^5}{I^2\omega^6}.
\label{eq:ger}
\end{equation}
The value of $\omega$ is expressed as
\begin{equation}
\omega=\frac{m_{p}c^2}{\hbar}.
\label{eq:ome}
\end{equation}
In view of (\ref{eq:imo})-(\ref{eq:ome}), the probability of
proton decay is of order
\begin{equation}
w\sim\frac{1}{m_{p}c^2}\frac{dE}{dt}=
\frac{Gm_{Pl}^{2}l_{Pl}^{4}m_{p}^{5}c^{5}}{\hbar^6}=
\frac{1}{t_{Pl}}\left(\frac{m_p}{m_{Pl}}\right)^5.
\label{eq:pdc}
\end{equation}
We obtain the probability of the proton decay in semiclassical
theory (\ref{eq:pdc}) the same as that in the quantum theory
(\ref{eq:pdq}).

At the Planck scale, radius of proton is equal to its Schwarzschild
radius and is equal to the Planck length
\begin{equation}
\frac{\hbar}{m_{Pl}c}=\frac{Gm_{Pl}}{c^2}=l_{Pl}.
\label{eq:rpr}
\end{equation}
From this the decay of proton into hypothetical fermions
prevents formation of the black holes, since at the Planck scale,
the probability of proton decay tends to unity. Hypothetical
fermions are candidates on the role of the dark matter.
Since hypothetical fermions
do not take part in electromagnetic, weak and strong
interactions, the dark matter consisting of hypothetical fermions
can be detected only by its gravitational potential.

\end{document}